\documentclass{article}
\usepackage{amsmath}
\usepackage{amsfonts}
\usepackage{amssymb}
\usepackage{graphicx}
\usepackage[english]{babel}
\usepackage[numbers,sort&compress]{natbib}
\usepackage{url}
\usepackage{doi}

\providecommand{\U}[1]{\protect\rule{.1in}{.1in}}

\begin{document}
\title{A real expectation value of the time-dependent\\non-Hermitian Hamiltonians$^{*}$}
\author{F. Kecita$^{a,b}$\thanks{E-mail: farouk.kecita@gmail.com \ }, A.
Bounames$^{a}$\thanks{E-mail: bounames@univ-jijel.dz \ } and M. Maamache$^{b}%
$\thanks{E-mail: maamache@univ-setif.dz \ }\\$^{(a)}$ {\small Laboratory of Theoretical Physics, University of Jijel, BP
98, Ouled Aissa, 18000 Jijel, Algeria.}\\$^{(b)}$ {\small Laboratoire de Physique Quantique et Syst\`{e}mes Dynamiques,
Facult\'{e} des Sciences, }\\{\small Universit\'{e} Ferhat Abbas S\'{e}tif 1, S\'{e}tif 19000, Algeria.}}
\date{}
\maketitle

\begin{abstract}
With the aim to solve the time-dependent Schr\"{o}dinger equation associated to 
a time-dependent non-Hermitian Hamiltonian, we introduce a unitary
transformation that maps the Hamiltonian to a time-independent $\mathcal{PT}%
$-symmetric one. Consequently, the solution of time-dependent Schr\"{o}dinger
equation becomes easily deduced and the evolution preserves the $\mathcal{C(}%
t\mathcal{)PT}$\emph{-}inner product, where\emph{ }$\mathcal{C(}t\mathcal{)}$
is a obtained from the charge conjugation operator $\mathcal{C}$ through
a time dependent unitary transformation. Moreover, the expectation value of
the non-Hermitian Hamiltonian in the\emph{ }$\mathcal{C(}t\mathcal{)PT}$\emph{
}normed states is guaranteed to be real. As an illustration, we present a
specific quantum system given by a quantum oscillator with time-dependent mass
subjected to a driving linear complex time-dependent potential.\newline
\\

PACS: 03.65.Ca, 03.65.-w \newline
\textbf{ Published in:}   Physica Scripta {\bf 96} (2021) 125265.  \doi{10.1088/1402-4896/ac3dbd}
\vspace{3cm}
\begin{flushright} {LPTh-Jijel-02/21} \end{flushright}

\end{abstract}

\vspace{2cm}
\textbf{$^{*}$This paper is dedicated to the memory of our friends and colleagues
Lahcene Alouani and Mebrouk Gherda died due to covid 19.}
\newpage

\section{Introduction}

It is commonly believed that the Hamiltonian must be Hermitian $H=H^{+}$ in
order to ensure that the energy spectrum (the eigenvalues of the Hamiltonian)
is real and that the time evolution of the theory is unitary (probability is
conserved in time), where the symbol `$+$' denotes the usual Dirac hermitian
conjugation; that is, transpose and complex conjugate. In 1998 this false
impression has been challenged by Bender and Boettcher \cite{Bender} who
showed numerically that a few one-dimensional quantum potentials $\ V(x)$ may
generate bound states $\psi(x)$ with real energies $E$ \ even when the
potentials themselves are not real. They show that because $\mathcal{PT}$-
symmetry is an alternative condition to Hermiticity. \ The central idea of
$\mathcal{PT}$ -symmetric quantum theory is to replace the condition that the
Hamiltonian of a quantum theory be Hermitian with the weaker condition: \emph{
}the invariance by space-time reflection. This allows one to construct and
study many new Hamiltonians that would previously have been ignored.

These two important discrete symmetry operators are parity $\mathcal{P}$ and
time reversal $\mathcal{T}$. The operators $\mathcal{P}$ and $\mathcal{T}$ are
defined by their effects on the dynamical variables $x$ and $p$. The operator
$\mathcal{P}$ is linear and has the effect of changing the sign of the
momentum operator $p$ and the position operator $x$ : $p$ $\rightarrow-p$ and
$x$ $\rightarrow-x$. The operator $\mathcal{T}$ is antilinear and has the
effect $p$ $\rightarrow-p$, $x$ $\rightarrow x$, and $i\rightarrow-i$. It is
crucial, of course, that when replacing the condition of Hermiticity by
$\mathcal{PT}$-symmetry, we preserve the key physical properties that a
quantum theory must have. We see that if the $\mathcal{PT}$- symmetry of the
Hamiltonian is not broken, then the Hamiltonian exhibits all of the features
of a quantum theory described by a Hermitian Hamiltonian. \ 

In order to have a coherent and unitary theory, Bender et al.
\cite{Bender2002} have defined the $\mathcal{PT}$ inner-product associated to
$\mathcal{PT}$ -symmetric Hamiltonians as follows
\begin{equation}
\left\langle f,g\right\rangle _{\mathcal{PT}}=\int_{C}dx\left[  \mathcal{PT}%
f(x)\right]  g(x),
\end{equation}
where $\mathcal{PT}f(x)=f^{\ast}(-x).$ The advantage of this inner product is
that the associated norm $(f,f),$ which independent of the global phase of
$f(x),$ is conserved in time. The application of this definition to the
eigenfunctions of $H$ and $\mathcal{PT}$ \ implies
\begin{equation}
\left\langle \psi_{m},\psi_{n}\right\rangle _{\mathcal{PT}}=(-1)^{n}%
\delta_{mn},\label{pt produit}%
\end{equation}

The situation here (that half of the the eigenfunctions of $H$ and
$\mathcal{PT}$ have positive norm and the other half have negative
norm) is analogous to the problem that Dirac encountered in formulating the
spinor wave equation in relativistic quantum theory. Following
Dirac,  Bender et al \cite{Bender2002}  constructed a linear operator
denoted by $\mathcal{C}$ and represented in position space as a sum
over the energy eigenstates of the Hamiltonian. The operator $\mathcal{C}$
is the observable that represents the measurement of the signature of
the $\mathcal{PT}$ norm of a state. The properties of the new
operator $\mathcal{C}$ resemble those of the charge conjugation
operator in quantum field theory.\textrm{ } Specifically, if the
energy eigenstates satisfy (\ref{pt produit}), then we have
$\mathcal{C}\psi_{n}=\left(  -1\right)  ^{n}\psi_{n}.$ $\mathcal{C}$,
called the charge conjugation symmetry with eigenvalues $\pm1,$
$\mathcal{C}^{2}=1$, such that $\mathcal{C}$ commutes with
the operator $\mathcal{PT}$ but not with the operators
$\mathcal{P}$ and $\mathcal{T}$ separately,  is the operator
observable that represents the measurement of the signature of the
$\mathcal{PT}$ norm of a state which determines its parity
type. We can regard $\mathcal{C}$ as representing the
operator that determines the $\mathcal{C}$ charge of the state. Quantum states having opposite $\mathcal{C}$ charge possess
opposite parity type. 

The introduction of operator $\mathcal{C}$ permits to formulate a positive $\mathcal{CPT}$ inner-product
\begin{equation}
\left\langle f,g\right\rangle _{\mathcal{CPT}}=\int_{C}dx\left[
\mathcal{CPT}f(x)\right]  g(x),
\end{equation}
thus Eq.(\ref{pt produit}) becomes
\begin{equation}
\left\langle \psi_{m},\psi_{n}\right\rangle _{\mathcal{CPT}}=\delta_{mn}\text{}.
\end{equation}

\textrm{The non-Hermitian }$\mathcal{PT}$\textrm{-symmetric models have been
successfully used for describing several physical systems such the plasmons in
nanoparticle systems \cite{nori1}, the problems related to the quantum
information theory \cite{nori2}, nonclassical light \cite{nori3} and the
stability of hydrogen molecules \cite{nori4}. }

\textrm{The generalization to} \textrm{ time-dependent non-Hermitian case have
been studied} \textrm{in}
\cite{5,Faria1,Faria2,most5,znojil2,znojil3,Bila,wang1,wang2,mus,fring1,fring2,khant,frith,luiz1,luiz2,mus2,Maamache2,mus3,koussa1,
koussa2, Mana1,Elaihar}. Note that the authors of Ref. \cite{Bender2011}
emphasize that in nonrelativistic quantum mechanics and in relativistic
quantum field theory, the time coordinate $t$ is a parameter and thus the
time-reversal operator $\mathcal{T}$ does not actually reverse the sign of
$t.$ Some authors adopt the fact that the operator $\mathcal{T}$ changes also
the sign of time\emph{\ }$t\rightarrow-t$\emph{\ }\cite{5, 4, 6, 7, 8,
9,10,11,Ramos,pedrosa,comment}, this case could lead sometimes to incorrect results.

\textrm{In this work,} we adopt the following strategy: we introduce a unitary
transformation $F(t)$ which commutes with the parity $\mathcal{P}$\ and maps
the solution $\left\vert \psi(t)\right\rangle $ of the time-dependent
Schr\"{o}dinger equation involving a non-Hermitian Hamiltonian $H(t)$ to the
solution $\left\vert \chi(t)\right\rangle $ involving a non Hermitian
Hamiltonian $\mathcal{H}$ required to be time-independent and $\mathcal{PT}%
$-symmetric. After performing this transformation, the problem becomes exactly
solvable and the evolution preserve the $\mathcal{CPT}$-scalar product
$\left\langle \chi(t)\right\vert \chi(t)\rangle_{\mathcal{CPT}}=\left\langle
\chi(t)\right\vert \mathcal{CP}\left\vert \chi(t)\right\rangle $. The other
essential ingredient of this theory is the construction of a positive-definite
inner product with respect to $H(t)$ being non self-adjoint, so that its
time-evolution operator is unitary and we obtain a consistent probabilistic
interpretation so that the Hamiltonian under study exhibits real mean values.
The most important step towards finding this positive-definite inner product
is thus to find a new operator, which we call $\mathcal{C(}t\mathcal{)}%
=F^{+}(t)\mathcal{C}F(t)$ such that, we obtain a conserved norm for our
original system described by the solution $\left\vert \psi(t)\right\rangle $
that is $\langle\psi(t)\left\vert \psi(t)\right\rangle _{\mathcal{C}%
(t)\mathcal{PT}}=\langle\chi(t)\left\vert \chi(t)\right\rangle _{\mathcal{CPT}%
}$, and the mean value of the time-dependent non-Hermitian Hamiltonian $H(t)$
is real in the new $\mathcal{C(}t\mathcal{)PT}$\emph{-}inner product. This is
the main result of this paper.

For this we introduce, in section\ 2, a formalism based on the time-dependent
unitary transformations is given in order to prove that the expectation value
of the time-dependent non-Hermitian Hamiltonian $H(t)$ is real in the new
$\mathcal{C(}t\mathcal{)PT}$\emph{-}inner product. In section\ 3, we
illustrate our formalism introduced in the previous section by treating a
non-Hermitian time-dependent quantum oscillator with time-dependent mass in
linear complex time-dependent potential. On the hand, in the Hermitian case
the time-dependent quantum harmonic have been extensively studied in the
literature in different ways
\cite{caldirola,kanai,abdalla,M2,M3,lopes,fan,ramos2}. Finally, section 4
concludes our work.

\section{Mean value of non-Hermitian time-dependent Hamiltonian}

Let us consider a non-hermitian time-dependent Hamiltonian $H(t)$ where the
quantum time evolution of the system is governed by the time-dependent
Schr\"{o}dinger equation (for simplicity we take$\ \hbar=1$)%
\begin{equation}
i\frac{\partial}{\partial t}\left\vert \psi(t)\right\rangle =H(t)\left\vert
\psi(t)\right\rangle .\text{ \ \ \ } \label{A5}%
\end{equation}

In order to study the evolution of the quantum systems associated to the
time-dependent Hamiltonian $H(t)$, we seek that this Hamiltonian can be
converted into a time-independent Hamiltonian by some time-dependent
transformations. To this end, we initially perform a unitary transformation
$F(t)$ on $\left\vert \psi(t)\right\rangle $%
\begin{equation}
\left\vert \chi(t)\right\rangle =F(t)\left\vert \psi(t)\right\rangle ,
\label{A6}%
\end{equation}
by inserting (\ref{A6}) in Eq. (\ref{A5}), we obtain the time dependent
Schr\"{o}dinger equation for the state $\left\vert \chi(t)\right\rangle $
\begin{equation}
i\frac{\partial}{\partial t}\left\vert \chi(t)\right\rangle =\mathcal{H}%
\left\vert \chi(t)\right\rangle , \label{A21}%
\end{equation}
such that the new Hamiltonian
\begin{equation}
\mathcal{H}=F(t)H(t)F^{+}(t)-iF(t)\frac{\partial F^{+}(t)}{\partial t},
\label{A8}%
\end{equation}
is time-independent and $\mathcal{PT}$-symmetric, i.e.;
\begin{equation}
\mathcal{H}\equiv\mathcal{H}_{0}^{\mathcal{PT}},
\end{equation}
\emph{ }\ its eigenstates $\left\vert \chi(t)\right\rangle $\emph{ }preserve
the $\mathcal{CPT}$-inner product\emph{ }%
\begin{equation}
\left\langle \chi(t)\right\vert \left.  \chi(t)\right\rangle _{\mathcal{CPT}%
}=\left\langle \chi(t)\right\vert \mathcal{CP}\left\vert \chi(t)\right\rangle
,
\end{equation}
and in this case the solution of the Schr\"{o}dinger equation (\ref{A21}) can
be written as%
\begin{equation}
\left\vert \chi(t)\right\rangle =\exp(-iEt)\left\vert \chi\right\rangle .
\end{equation}
where $\left\vert \chi\right\rangle $ is an eigenstate of \ $\mathcal{H}%
_{0}^{\mathcal{PT}}$ .

Knowing that our interest is the mean value of the non-Hermitian Hamiltonian
$H(t)$, for this aim we calculate firstly the expectation value of the
Hamiltonian $\mathcal{H}_{0}^{\mathcal{PT}}$%
\begin{equation}
\left\langle \mathcal{H}_{0}^{\mathcal{PT}}\right\rangle _{\mathcal{CPT}%
}=\left\langle \chi(t)\right\vert \mathcal{CPH}_{0}^{\mathcal{PT}}\left\vert
\chi(t)\right\rangle =\left\langle \chi(t)\right\vert \mathcal{CP}\left[
FH(t)F^{+}-iF\frac{\partial F^{+}}{\partial t}\right]  \left\vert
\chi(t)\right\rangle ,
\end{equation}
from which we deduce that is
\begin{equation}
\left\langle \chi(t)\right\vert \mathcal{CP}\left[  FH(t)F^{+}\right]
\left\vert \chi(t)\right\rangle =\left\langle \mathcal{H}_{0}^{\mathcal{PT}%
}\right\rangle _{\mathcal{CPT}}+\left\langle \chi(t)\right\vert \mathcal{CP}%
\left[  iF\frac{\partial F^{+}}{\partial t}\right]  \left\vert \chi
(t)\right\rangle ,
\end{equation}
we note that the first term is nothing other than\emph{ } the expectation
value of the Hamiltonian $H(t)$ with a new $\mathcal{C}(t)\mathcal{PT}$-inner
product
\begin{equation}
\left\langle \chi(t)\right\vert \mathcal{CP}\left[  FH(t)F^{+}\right]
\left\vert \chi(t)\right\rangle =\left\langle \psi(t)\right\vert
\mathcal{C}(t)\mathcal{P}H(t)\left\vert \psi(t)\right\rangle =\left\langle
H(t)\right\rangle _{\mathcal{C}(t)\mathcal{PT}}\text{ },
\end{equation}
where $[\mathcal{P},F(t)]=0$ and the new operator $C(t)$ is defined as
$C(t)=F^{+}(t)CF(t)$, which is similar to the operator $C$ in the sense that
verifies the property $\mathcal{C}^{2}\mathcal{(}t\mathcal{)}=1$ since
$\mathcal{C}^{2}=1$.

Finally,%
\begin{equation}
\left\langle H(t)\right\rangle _{\mathcal{C}(t)\mathcal{PT}}=\left\langle
\mathcal{H}_{0}^{\mathcal{PT}}\right\rangle _{\mathcal{CPT}}+\left\langle
\chi(t)\right\vert \mathcal{CP}\left[  iF\frac{\partial F^{+}}{\partial
t}\right]  \left\vert \chi(t)\right\rangle .
\end{equation}

Indeed, since $\mathcal{H}_{0}^{\mathcal{PT}}$ \ is\emph{ }$\mathcal{PT}%
$\emph{ }symmetric and $F$\emph{\ }is unitary$,$ the expectation value
$\left\langle H(t)\right\rangle _{\mathcal{C}(t)\mathcal{PT}}$\emph{ }is
guaranteed to be real. To our knowledge, this general result is new for
explicitly time-dependent non-Hermitian systems.

\section{Application: non-Hermitian time-dependent mass forced oscillators}

Let us consider\ a class of one dimensional time-dependent harmonic
oscillators with variable mass $m(t)=m_{0}\alpha(t)$ subjected to a driving
linear complex time-dependent potential, in the form $i\lambda(t)x$, described
by the following non-Hermitian Hamiltonian%
\begin{equation}
H(t)=\frac{p^{2}}{2m_{0}\alpha(t)}+\alpha(t)\frac{m_{0}\omega^{2}(t)}{2}%
x^{2}+ix\sqrt{\alpha(t)}, \label{B3}%
\end{equation}
where $\alpha(t)$ is a positive real time-dependent function, $x$ and $p$ are
the canonical conjugates position and momentum operators satisfying $\left[
x,p\right]  =i$. The function $\lambda(t)$ in the complex potential has been
choosen $\lambda(t)=\sqrt{\alpha(t)}$ in order to obtain in Eq. (\ref{A8}) a
time-independent $\mathcal{PT}$-symmetric Hamiltonian $\mathcal{H}%
_{0}^{\mathcal{PT}}.$Without loss of generalities, we choose $\omega
(t)=\omega$ as a constant. The mass $m_{0}$ and the frequency $\omega$ are the
characteristic parameters of the quantum system.

We show that the exact solution of the time-dependent Schr\"{o}dinger equation
(\ref{A5}) can be found by introducing two consecutive unitary
transformations. In order to solve the Schr\"{o}dinger with the Hamiltonian
specified by (\ref{B3}), we first try to eliminate the time-dependent
parameter $\alpha(t)$\textsl{. }This can be achieved by the transformation

\begin{equation}
F_{1}(t)=\exp\left[  -\frac{i}{2}\left\{  x,p\right\}  \ln\left(  \sqrt
{\alpha(t)}\right)  \right]
\end{equation}

The unitary operator $F_{1}(t)$ has the properties\textsl{ }%
\begin{equation}
F_{1}xF_{1}^{+}=\frac{x}{\sqrt{\alpha(t)}},\ \ \ \ F_{1}pF_{1}^{+}%
=p\sqrt{\alpha(t)}\ ,
\end{equation}

In a representation $x$, the wave function is given by\textsl{ }%
\begin{equation}
\left\langle x\right\vert F_{1}\left\vert \phi\right\rangle =\alpha^{-\frac
{1}{2}}\phi\left(  x\alpha^{-\frac{1}{2}}\right)  .
\end{equation}

Suppose that\textsl{ }%
\begin{equation}
\left\vert \phi(t)\right\rangle =F_{1}(t)\left\vert \psi(t)\right\rangle ,
\label{f1}
\end{equation}

Substituting (\ref{f1}) into (\ref{A5})ruled by the Hamiltonian (\ref{B3}), we
find the equation of motion for\textsl{ }$\left\vert \phi(t)\right\rangle $%

\begin{equation}
i\frac{\partial}{\partial t}\left\vert \phi(t)\right\rangle =H_{1}%
(t)\left\vert \phi(t)\right\rangle , \label{f2}%
\end{equation}
where the Hamiltonian
\begin{align}
H_{1}(t)  &  =F_{1}(t)H(t)F_{1}^{+}(t)-iF_{1}(t)\frac{\partial F_{1}^{+}%
(t)}{\partial t}\\
&  =\frac{p^{2}}{2m_{0}}+\frac{m_{0}\omega^{2}}{2}x^{2}+ix+\frac{1}{4}%
\frac{\dot{\alpha}(t)}{\alpha(t)}(xp+px) \label{f3}%
\end{align}
look like the time-independent harmonic oscillators with variable mass $m_{0}$
subjected to a driving linear complex time-independent potential plus a time
dependent\textsl{ }$(xp+px)$\textsl{ }terms. In order to obtain the usual
time-dependent harmonic oscillator with a perturbative linear potential, we
remove the cross term in (\ref{f3}) via the transformation\textsl{ }%
\begin{equation}
F_{2}(t)=\exp\left[  i\frac{m_{0}\dot{\alpha}(t)}{4\alpha(t)}x^{2}\right]
,\text{ } \label{f4}%
\end{equation}
\textsl{ }where its properties are\textsl{ }%
\begin{equation}
F_{2}xF_{2}^{+}=x,\ \ \ \ F_{2}pF_{2}^{+}=-\frac{m_{0}\dot{\alpha}(t)}%
{2\alpha(t)}x,
\end{equation}

Thus, the following unitary transformation $F(t)=F_{2}(t)F_{1}(t)$
\begin{equation}
F(t)=\exp\left[  i\frac{m_{0}\dot{\alpha}(t)}{4\alpha(t)}x^{2}\right]
\exp\left[  -\frac{i}{2}\left\{  x,p\right\}  \ln\left(  \sqrt{\alpha
(t)}\right)  \right]  ,\text{ } \label{AA1}%
\end{equation}
transforms the canonical operators $x$ and $p$ and their squares $x^{2}$ and
$p^{2}$ as follows
\[
FxF^{+}=\frac{x}{\sqrt{\alpha(t)}},\ \ \ \ FpF^{+}=p\sqrt{\alpha(t)}%
-\frac{m_{0}\dot{\alpha}(t)}{2\sqrt{\alpha(t)}}x\ ,
\]%
\begin{equation}
Fp^{2}F^{+}=\alpha(t)x^{2}-\frac{1}{2}m_{0}\dot{\alpha}(t)\left\{
x,p\right\}  +\frac{m_{0}^{2}\dot{\alpha}^{2}(t)}{4\alpha(t)}x^{2},\text{
\ \ }Fx^{2}F^{+}=\frac{x^{2}}{\alpha(t)},\text{\ } \label{A11}%
\end{equation}
\textrm{therefore, the transformed Hamiltonian}\emph{ }\textrm{(\ref{A8})
reads}
\begin{equation}
\mathcal{H}=\frac{p^{2}}{2m_{0}}+\frac{1}{2}m_{0}{\Omega}^{2}x^{2}+ix.
\label{R12}%
\end{equation}
\textrm{where }%
\begin{equation}
{\Omega}^{2}{=}\left(  \omega^{2}+\frac{1}{4}\frac{\dot{\alpha}^{2}(t)}%
{\alpha^{2}(t)}-\frac{\ddot{\alpha}(t)}{2\alpha(t)}\right)  \label{2}%
\end{equation}

\textrm{The central idea in this procedure is to require that the Hamiltonian
(\ref{R12}) governing the evolution of}\emph{ }$\left\vert \chi
(t)\right\rangle $ \textrm{is time-independent. This is achieved by setting
the global time-dependent frequency appearing in \ (\ref{R12}) equal to a real
constant denoted by} ${\Omega}_{0}^{2}$\ \textrm{so that its time-derivatve
leads to an auxiliary equation of the form}
\begin{equation}
\ddot{\alpha}-\frac{\dot{\alpha}^{2}}{2\alpha}+2\alpha\left(  {\Omega}_{0}%
^{2}-\omega^{2}\right)  =0, \label{aux}%
\end{equation}
\textrm{the resulting time independent non-Hermitian Hamiltonian}
\begin{equation}
\mathcal{H}_{0}^{\mathcal{PT}}=\frac{p^{2}}{2m_{0}}+\frac{1}{2}m_{0}\Omega
_{0}^{2}x^{2}+ix, \label{B1}%
\end{equation}
\textrm{is }$\mathcal{PT}$\textrm{-symmetric.}

Note that when taking\emph{ }$\alpha(t)=\frac{1}{\rho^{2}(t)},$\emph{ }the
above auxiliary equation (\ref{aux}) is transformed to the following new
auxiliary equation%
\begin{equation}
\overset{..}{\rho}+\left(  {\Omega}_{0}^{2}\mathcal{-}\omega^{2}\right)
\rho=0.
\end{equation}
which admits the following solutions:

$\bullet$ for ${\Omega}_{0}^{2}\mathcal{>}\omega^{2}:$ $\rho(t)=$
$A\exp\left(  it\sqrt{{\Omega}_{0}^{2}\mathcal{-}\omega^{2}}\right)
+B\exp\left(  -it\sqrt{{\Omega}_{0}^{2}\mathcal{-}\omega^{2}}\right)  .$ For
an appropriate choice of the constants: $A=B,$ we obtain the expression of
$\alpha(t)$\ as $\alpha(t)=\frac{1}{A^{2}\cos^{2}\left(  t\sqrt{{\Omega}%
_{0}^{2}\mathcal{-}\omega^{2}}\right)  }$ \emph{.}

$\bullet$ for ${\Omega}_{0}^{2}\mathcal{<}\omega^{2}:$ $\rho(t)=$
$A\exp\left(  t\sqrt{\omega^{2}-{\Omega}_{0}^{2}}\right)  +B\exp\left(
-t\sqrt{\omega^{2}-{\Omega}_{0}^{2}}\right)  .$ For an appropriate choice of
the constants: $A=B,$ we obtain the expression of $\alpha(t)$\ as
$\alpha(t)=\frac{1}{A^{2}\cosh^{2}\left(  t\sqrt{\omega^{2}-{\Omega}_{0}^{2}%
}\right)  },$ and when $B=0$ and $A\neq0$ the expression of $\alpha(t)$ is
$\alpha(t)=\frac{1}{A^{2}}\exp\left(  -2t\sqrt{\omega^{2}-{\Omega}_{0}^{2}%
}\right)  $ and the Hamiltonian $H(t)$ corresponds to the Caldirola-Kanai
oscillator \cite{caldirola,kanai}.

\subsection{Analysis of the expectation value of the Hamiltonian}

The eigenequation of the $\mathcal{PT}$ -symmetric Hamiltonian $\mathcal{H}%
_{0}^{\mathcal{PT}}$ has the form%
\begin{equation}
\mathcal{H}_{0}^{\mathcal{PT}}\left\vert \chi_{n}(x)\right\rangle
=E_{n}\left\vert \chi_{n}(x)\right\rangle ,
\end{equation}
and the solution of the corresponding Schr\"{o}dinger equation (\ref{A21}) can
be written as%
\begin{equation}
\left\vert \chi_{n}(x,t)\right\rangle =\exp(-iE_{n}t)\left\vert \chi
_{n}(x)\right\rangle .
\end{equation}

Let us introduce a non unitary transformation of the form%
\begin{equation}
U=\exp\left[  -\frac{p}{m_{0}\Omega_{0}^{2}}\right]  ,
\end{equation}
such that
\begin{equation}
\left\vert \chi_{n}(x)\right\rangle =U\left\vert \varphi_{n}(x)\right\rangle .
\end{equation}

The action of $U$ maps the\emph{ }$\mathcal{PT}$-symmetric Hamiltonian
$H_{0}^{\mathcal{PT}}$ to a Hermitian one as
\begin{equation}
h=U^{-1}\mathcal{H}_{0}^{\mathcal{PT}}U=\frac{p^{2}}{2m_{0}}+\frac{m_{0}%
\Omega_{0}^{2}}{2}x^{2}-\frac{1}{2m_{0}\Omega_{0}^{2}},\label{B14}%
\end{equation}
where the eigenfunctions $\left\vert \varphi_{n}(x)\right\rangle $ of the
Hermitian Hamiltonian $h$ are
\begin{equation}
\left\vert \varphi_{n}(x)\right\rangle =\left[  \frac{\sqrt{m_{0}\Omega_{0}}%
}{n!2^{n}\sqrt{\pi\hbar}}\right]  ^{1/2}\exp\left(  -\frac{m_{0}\Omega_{0}%
}{2\hbar}x^{2}\right)  H_{n}\left[  x\left(  \frac{m_{0}\Omega_{0}}{\hbar
}\right)  ^{1/2}\right]  .\label{B11}%
\end{equation}
Then, the solutions $\left\vert \chi_{n}(x,t)\right\rangle $ are obtained as%
\begin{align}
\left\vert \chi_{n}(x,t)\right\rangle  &  =\exp(-iE_{n}t)U\left\vert
\varphi_{n}(x)\right\rangle ,\nonumber\\
\left\vert \chi_{n}(x,t)\right\rangle  &  =\left[  \frac{\sqrt{m_{0}\Omega
_{0}}}{n!2^{n}\sqrt{\pi\hbar}}\right]  ^{1/2}\exp(-iE_{n}t)\exp\left[
-\frac{p}{m_{0}\Omega_{0}^{2}}\right]  \exp\left(  -\frac{m_{0}\Omega_{0}%
}{2\hbar}x^{2}\right)  H_{n}\left[  \left(  \frac{m_{0}\Omega_{0}}{\hbar
}\right)  ^{1/2}x\right]  ,
\end{align}
where the eigenvalues%
\begin{equation}
E_{n}=\hbar\Omega_{0}\left(  n+\frac{1}{2}\right)  -\frac{1}{2m_{0}\Omega
_{0}^{2}},
\end{equation}
are real and $H_{n}$ is the Hermite polynomial of order $n$. 

A more general way to represent the $\mathcal{C}$
operator is to express it generically in terms of the fundamental
dynamical operators $x$ and $p$ $:$ $\mathcal{C=}e^{Q(x,p)}.$ The
exact formula of $\mathcal{C}$ associated to the theory
described by the Hamiltonian (\ref{B1}) is given as  a
function of the parity operator $\mathcal{P}$ as
\begin{equation}
\mathcal{C}=\exp\left[  \frac{2}{m_{0}\Omega_{0}^{2}}p\right]  \mathcal{P}%
,\label{B5}%
\end{equation}
such that the operator \textrm{ }$\mathcal{C}$ commute
with\textrm{ }$\mathcal{PT}$ and $\mathcal{H}_{0}^{\mathcal{PT}}$ ,i.e.,
$[\mathcal{C},\mathcal{PT}]=[\mathcal{C},\mathcal{H}_{0}^{\mathcal{PT}}]=0.$

We can easily verify that the $\mathcal{CPT}$-inner product is conserved%
\begin{equation}
\left\langle \chi_{_{n}}(x,t)\right.  \left\vert \chi_{_{n}}(x,t)\right\rangle
_{\mathcal{CPT}}=\left\langle \chi_{_{n}}(x)\right\vert \mathcal{CP}\left\vert
\chi_{_{n}}(x)\right\rangle =\left\langle \varphi_{n}\right\vert
U\mathcal{CP}U\left\vert \varphi_{n}\right\rangle =\left\langle \varphi
_{n}(x)\right.  \left\vert \varphi_{n}(x)\right\rangle =1.\label{B6}%
\end{equation}

Now it is not difficult to calculate the expectation value of the Hamiltonian
$\left\langle H(t)\right\rangle _{\mathcal{C}(t)\mathcal{PT}}$ defined
previously%
\begin{equation}
\left\langle H(t)\right\rangle _{\mathcal{C}(t)\mathcal{PT}}=E_{n}-\frac
{\dot{\alpha}(t)}{4\alpha(t)}\left\langle \varphi_{n}(x)\right\vert
U^{-1}\left\{  x,p\right\}  U\left\vert \varphi_{n}(x)\right\rangle +\left(
\frac{m_{0}\ddot{\alpha}(t)}{4\alpha(t)}\right)  \left\langle \chi_{_{n}%
}(x)\right\vert \mathcal{CP}x^{2}\left\vert \chi_{_{n}}(x)\right\rangle ,
\end{equation}
thus%
\begin{align}
\left\langle H(t)\right\rangle _{\mathcal{C}(t)\mathcal{PT}}  &  =E_{n}%
-\frac{\dot{\alpha}(t)}{4\alpha(t)}\left\langle \varphi_{n}(x)\right\vert
\left\{  x,p\right\}  \left\vert \varphi_{n}(x)\right\rangle \nonumber\\
&  +\frac{\dot{\alpha}(t)}{2\alpha(t)}\frac{i}{m_{0}\Omega_{0}^{2}%
}\left\langle \varphi_{n}(x)\right\vert p\left\vert \varphi_{n}%
(x)\right\rangle +\left(  \frac{m_{0}\ddot{\alpha}(t)}{4}\right)  \left\langle
x^{2}\right\rangle _{\mathcal{CPT}},
\end{align}
where $\left\langle x^{2}\right\rangle _{\mathcal{CPT}}=\left\langle
\chi_{_{n}}(x)\right\vert \mathcal{CP}x^{2}\left\vert \chi_{_{n}%
}(x)\right\rangle .$ By using the following relations
\begin{align}
\left\langle \varphi_{n}(x)\right\vert x\left\vert \varphi_{n}%
(x)\right\rangle  &  =\left\langle \varphi_{n}(x)\right\vert p\left\vert
\varphi_{n}(x)\right\rangle =0,\\
\left\langle \varphi_{n}(x)\right\vert x^{2}\left\vert \varphi_{n}%
(x)\right\rangle  &  =\frac{\hbar}{m_{0}\Omega_{0}}\left(  n+\frac{1}%
{2}\right)  ,\\
\left\langle \varphi_{n}(x)\right\vert p^{2}\left\vert \varphi_{n}%
(x)\right\rangle  &  =m_{0}\Omega_{0}\hbar\left(  n+\frac{1}{2}\right)  ,
\end{align}%
\begin{equation}
\left\langle \varphi_{n}(x)\right\vert \left\{  x,p\right\}  \left\vert
\varphi_{n}(x)\right\rangle =0,
\end{equation}
and%
\begin{equation}
\left\langle x^{2}\right\rangle _{\mathcal{CPT}}=\frac{\hbar}{m_{0}\Omega
}\left(  n+\frac{1}{2}\right)  -\frac{1}{\left(  m_{0}\Omega_{0}^{2}\right)
^{2}},
\end{equation}
we get the expectation value of $H(t)$ as%
\begin{equation}
\left\langle H(t)\right\rangle _{\mathcal{C}(t)\mathcal{PT}}=E_{n}+\left(
\frac{m_{0}\ddot{\alpha}(t)}{4\alpha(t)}\right)  <x^{2}>_{\mathcal{CPT}}%
=E_{n}+\frac{\ddot{\alpha}(t)}{4\alpha(t)}\left[  \frac{\hbar}{\Omega_{0}%
}\left(  n+\frac{1}{2}\right)  -\frac{1}{m_{0}\Omega_{0}^{4}}\right]  ,
\end{equation}
which is real for any positive real time-dependent function $\alpha(t)$ and
more simple than the result given in Eq. (28) in Ref. \cite{5} with less
constraints on the parameters of the problem.

\subsection{Uncertainty relation and probability density}

Now, we calculate\ the expectation values $\left\langle x\right\rangle
_{\mathcal{C}(t)\mathcal{PT}}$ , $\left\langle x^{2}\right\rangle
_{\mathcal{C}(t)\mathcal{PT}},$ $\left\langle p\right\rangle _{\mathcal{C}%
(t)\mathcal{PT}}$ and $\left\langle p^{2}\right\rangle _{\mathcal{C}%
(t)\mathcal{PT}}$ in the states $\psi_{n}(x,t)$ of $H(t)$ defined in
Eq.(\ref{B3}). In the same way, using the $\mathcal{CPT}$-inner product
(\ref{B6}) and after straightforward calculation we obtain that%
\begin{equation}
\left\langle x\right\rangle _{\mathcal{C}(t)\mathcal{PT}}=\left\langle
\psi_{n}(x,t)\right\vert F^{+}\mathcal{CP}Fx\left\vert \psi_{n}%
(x,t)\right\rangle =-\frac{i}{m_{0}\Omega_{0}^{2}}\frac{1}{\sqrt{\alpha(t)}},
\end{equation}

\begin{equation}
\left\langle x^{2}\right\rangle _{\mathcal{C}(t)\mathcal{PT}}=\left\langle
\psi_{n}(x,t)\right\vert F^{+}\mathcal{CP}Fx^{2}\left\vert \psi_{n}%
(x,t)\right\rangle =\left(  n+\frac{1}{2}\right)  \frac{\hbar}{m_{0}\Omega
_{0}\alpha(t)}-\frac{1}{\alpha(t)\left(  m_{0}\Omega_{0}^{2}\right)  ^{2}},
\end{equation}%
\begin{equation}
\left\langle p\right\rangle _{\mathcal{C}(t)\mathcal{PT}}=\left\langle
\psi_{n}(x,t)\right\vert F^{+}\mathcal{CP}Fp\left\vert \psi_{n}%
(x,t)\right\rangle =\frac{i}{2\Omega_{0}^{2}}\frac{\dot{\alpha}(t)}%
{\sqrt{\alpha(t)}},
\end{equation}%
\begin{align}
\left\langle p^{2}\right\rangle _{\mathcal{C}(t)\mathcal{PT}}  &
=\left\langle \psi_{n}(x,t)\right\vert F^{+}\mathcal{CP}Fp^{2}\left\vert
\psi_{n}(x,t)\right\rangle ,\nonumber\\
\left\langle p^{2}\right\rangle _{\mathcal{C}(t)\mathcal{PT}}  &  =\hbar
\Omega_{0}\left(  n+\frac{1}{2}\right)  m_{0}\alpha(t)+\left(  \frac{m_{0}%
\dot{\alpha}(t)}{2}\right)  ^{2}\left[  \left(  n+\frac{1}{2}\right)
\frac{\hbar}{m_{0}\Omega_{0}\alpha(t)}-\frac{1}{\alpha(t)\left(  m_{0}%
\Omega_{0}^{2}\right)  ^{2}}\right]  .
\end{align}

We calculate also the position and momentum uncertainties
\begin{equation}
\Delta x=\sqrt{\left\langle x^{2}\right\rangle _{\mathcal{C}(t)\mathcal{PT}%
}-\left\langle x\right\rangle _{\mathcal{C}(t)\mathcal{PT}}^{2}}=\left[
\frac{\hbar}{m_{0}\Omega_{0}\alpha(t)}\left(  n+\frac{1}{2}\right)  \right]
^{1/2},
\end{equation}%
\begin{equation}
\Delta p=\sqrt{\left\langle p^{2}\right\rangle _{\mathcal{C}(t)\mathcal{PT}%
}-\left\langle p\right\rangle _{\mathcal{C}(t)\mathcal{PT}}^{2}}=\frac
{1}{\Delta x}\left[  \left(  n+\frac{1}{2}\right)  ^{2}+\left(  \frac
{m_{0}\dot{\alpha}(t)}{2}\right)  ^{2}\Delta x^{4}\right]  ^{1/2}.
\end{equation}

Thus, the uncertainty product is given by
\begin{equation}
\Delta x\Delta p=\left(  n+\frac{1}{2}\right)  \sqrt{1+\left(  \frac{\hbar
\dot{\alpha}(t)}{2\Omega_{0}\alpha(t)}\right)  ^{2}},\label{A50.}%
\end{equation}
it is easy to check that the uncertainty product (\ref{A50.}) is always real
and greater than or equal to $\frac{1}{2}$ and, consequently, it is physically
acceptable for any value of $n$. The uncertainty product takes the minimal
value $\Delta x\Delta p=\frac{1}{2}$ only for $n=0$ and $\alpha(t)=$constant,
i.e., for time independent mass oscillators.

Finally, the probability density of the wavefunction %
$\psi_{n}(x,t)$ of $H(t)$ is in the form
\begin{equation}
\left\vert U^{-1}F\psi_{n}(x,t)\right\vert ^{2}=\left\vert U^{-1}\chi_{_{n}%
}(x,t)\right\vert ^{2}=\left\vert \varphi(x)\right\vert ^{2}=\varphi_{n}%
^{\ast}(x)\varphi_{n}(x),
\end{equation}
thus%
\begin{equation}
\left\vert U^{-1}F\psi_{n}(x,t)\right\vert ^{2}=\left[  \frac{\sqrt
{m_{0}\Omega_{0}}}{n!2^{n}\sqrt{\pi\hbar}}\right]  \exp\left(  -\frac
{m_{0}\Omega_{0}}{\hbar}x^{2}\right)  \left(  H_{n}\left[  \left(  \frac
{m_{0}\Omega_{0}}{\hbar}\right)  ^{1/2}x\right]  \right)  ^{2},\label{A56}%
\end{equation}
is the same as the probability density of the eigenstate $\chi_{_{n}
}(x,t)$ of time independent $\mathcal{H}_{0}^{\mathcal{PT}}$ which is also
equal to the probability density of the eigenstate $\varphi_{n}(x)$
(\ref{B11}) of the standard harmonic oscillator (\ref{B14}). Clearly,
$\varphi_{n}(x)$ are elements from $L^{2}(R)$,
and therefore the condition (\ref{A56}) yields that

\begin{equation}
\int\left\vert \varphi_{n}(x)\right\vert ^{2}dx=\left[  \frac{\sqrt
{m_{0}\Omega_{0}}}{n!2^{n}\sqrt{\pi\hbar}}\right]  ^{1/2}%
{\displaystyle\int}
\exp\left(  -\frac{m_{0}\Omega_{0}}{\hbar}x^{2}\right)  \left(  H_{n}\left[
x\left(  \frac{m_{0}\Omega_{0}}{\hbar}\right)  ^{1/2}\right]  \right)
^{2}dx=1\label{r11}%
\end{equation}
under this observation, we deduce that the probability is finite.

\section{Conclusion}

The essential ingredient of quantum mechanical non Hermitian theory is
the construction of a positive-definite inner product, so that its probability
is conserved in time. The operator $\mathcal{C(}t\mathcal{)}=F^{+}%
(t)\mathcal{C}F(t)$ confer to the norm its conservation. The main result of this paper is that the mean value of a time-dependent non-Hermitian Hamiltonian $H(t)$ is real in the new
$\mathcal{C(}t\mathcal{)PT}$\emph{-}inner product. For this, we introduced a
unitary transformation $F(t)$ that reduces the study of time-dependent
non-Hermitian Hamiltonian $H(t)$ to the study of time-independent
$\mathcal{PT}$-symmetric Hamiltonian $\mathcal{H}_{0}^{\mathcal{PT}}$, and
derived the analytical solution of the Schr\"{o}dinger equation of the initial
system. Then, we defined a new $\mathcal{C(}t\mathcal{)PT}$-inner product and
showed that the evolution preserves it. Furthermore, we proved that the
expectation value of the time-dependent non-Hermitian Hamiltonian $H(t)$ is
real in the $\mathcal{C(}t\mathcal{)PT}$\emph{ }normed states since the
transformation $F(t)$ is unitary and $[\mathcal{P},F(t)]=0$. As an
illustration, we have investigated a class of quantum time-dependent mass
oscillators with a complex linear driving force. The\ expectation value of the
Hamiltonian, the uncertainty relation and probability density have been also
calculated. 

\paragraph{Acknowledgments}

One of the authors (M.M) would like to thank Dr. W. Koussa and Dr. N. Mana for
their helpful contribution.

\end{document}